\documentclass[12pt]{iopart}
\usepackage{amsfonts}

\usepackage{graphicx}

\begin{document}

\title[Induced vacuum energy-momentum tensor]{Induced vacuum energy-momentum tensor in \\ the background of a cosmic string}

\author{Yu A Sitenko$^1$ and N D Vlasii$^{1,2}$}

\address{$^{1}$ Bogolyubov Institute for Theoretical Physics,
National Academy of Sciences, Kyiv, 03680, Ukraine}
\address{$^{2}$ Physics Department, Taras Shevchenko National University of
Kyiv, \\
Kyiv, 01601, Ukraine}

\ead{yusitenko@bitp.kiev.ua}

\begin{abstract}
A massive scalar field is quantized in the background of a cosmic
string which is generalized to a static flux-carrying codimension-2
brane in the locally flat multidimensional space-time. We find that
the finite energy-momentum tensor is induced in the vacuum. The
dependence of the tensor components on the brane flux and tension,
as well as on the coupling to the space-time curvature scalar, is
comprehensively analyzed. The tensor components are holomorphic
functions of space dimension, decreasing exponentially with the
distance from the brane. The case of the massless quantized scalar
field is also considered, and the relevance of Bernoulli's
polynomials of even order for this case is discussed.
\end{abstract}

\pacs{11.15.-q, 04.50.-h, 11.27.+d, 98.80.Cq}


{\it Keywords}: vacuum polarization, vortex, multidimensional
conical space-time

\submitto{\CQG}

\section{Introduction}

String-like configurations of classical background fields can
polarize the vacuum of quantized matter fields, resulting in the
emergence of a nonzero vacuum expectation value of the
energy-momentum tensor in spatial regions where the classical field
is zero. A particular example of a string-like configuration is a
tube of the magnetic flux lines, that is formed inside a
current-carrying long and thin solenoid. But a more general
string-like structure is the Abrikosov-Nielsen-Olesen vortex
\cite{Abri,Nie} arising as a topological defect in the aftermath of
a phase transition with spontaneous breakdown of a continuous
symmetry; the condition of its arising is that the first homotopy
group of the group space of the broken symmetry group be nontrivial.
The vortex, apart from its flux, is characterized by nonzero energy
distributed along its axis, which in its turn, according to general
relativity, is a source of gravity. It can be shown \cite{Vi,Go,Ga}
that this source makes space outside the vortex to be conical with
the deficit angle related to the energy density per unit length of
the vortex. Since the squared Planck length enters as a factor
before the stress-energy tensor in the Einstein-Hilbert equation,
the deviations from the Minkowskian metric are of the order of
squared quotient of the Planck length to the correlation length, the
latter characterizing the thickness of the vortex. For
superconductors this quotient is vanishingly small and effects of
conicity are surely negligible \cite{Hub}; hence, the appropriate
vortices known as the Abrikosov ones are much the same as the
solenoidal magnetic vortices (with one important distinction that
the flux of the former is quantized). However, topological defects
of the vortex type arise in cosmology under the name of cosmic
strings \cite{Vil, Hin}. Cosmic strings with the thickness of the
order of the Planck length are definitely ruled out by astrophysical
observations, and there remains a room for cosmic strings with the
thickness which is more than 3.5 orders larger than the Planck
length, that corresponds to the deficit angles bounded from above by
the value of $2.5\times 10^{-6}$ radians (see \cite{Bat}). Cosmic
strings associated with spontaneous breakdown of global symmetries
are called global (or axionic) strings; they are characterized by
zero flux. Cosmic strings associated with breakdown of gauge
symmetries are called gauge (or local) strings; they are
characterized by nonzero flux.

Cosmic strings can have various astrophysical effects, in
particular, they serve as plausible sources of detectable
gravitational waves \cite{Dam,Bran,JaSi}, gamma-ray bursts
\cite{Ber}, high-energy cosmic rays \cite{Bra}. Although the primary
role of cosmic strings in the formation of the large-scale structure
of the Universe is ruled out by observation of cosmic microwave
background by COBE and WMAP \cite{Bev,Pog}, cosmic strings can
produce a mechanism for the generation of the primordial magnetic
field in the early Universe \cite{DaD,Gwy,Si9}. The interest to
cosmic strings is augmented in the last decade owing to the finding
that they emerge at the end of inflation in the framework of
superstring models \cite{Sar,Jean} (see also reviews
\cite{Pol,Dav,Sak,Cop}). The details of the cosmic string formation
mechanism can be quite different, either fundamental (super)strings
or Dirichlet branes are stretched to a macroscopic size, but such
structures arize almost inevitably in supergravity models with large
extra dimensions \cite{Ark}. In this respect it is appropriate to
consider a generalization of a cosmic string in 4-dimensional
space-time to a codimension-2 brane in higher-dimensional
space-time. Such a brane appears also in the construction of
brane-world cosmologies with two extra dimensions (see
\cite{Ghe,Lee,Bur} and references therein), as well as in the study
of quantum aspects of the black hole physics, playing a key role in
the formation of chains of Kaluza-Klein bubbles with black holes
\cite{Elv,Kas}. Therefore, the properties of the vacuum of quantized
matter in the background of a codimension-2 brane deserve
investigation.

A study of the vacuum polarization effects in the cosmic string
background has a long history. The induced vacuum energy-momentum
tensor was considered for the cases of global \cite{Deu,Hel,Lin} and
gauge \cite{Fro,Dow} strings; in the above the quantized fields were
assumed to be massless, whereas the results for the nonvanishing
mass were either incomplete or of restricted use, see
\cite{Gui,Mor,Iel}. The effects of nonvanishing mass were
exhaustively studied for the case of a negligible deficit angle only
\cite{Si3} (see also \cite{Si8,SiB}). In this respect it should be
noted that the induced vacuum current in the cosmic string
background is known both in the massive \cite{Si9} and massless
\cite{Sri} cases.

The aim of the present paper is to obtain  the energy-momentum
tensor which is induced in the vacuum of the quantized massive
scalar field in the cosmic string background. In the next section,
the vacuum expectation value of the energy-momentum tensor is in
general defined with the use of the point-splitting regularization.
Green's function in the cosmic string background is obtained in
section 3. The induced vacuum energy-momentum tensor is obtained in
section 4. Some limiting cases of our result are discussed in
section 5, while the case of the massless quantized scalar field is
considered in section 6. The results are summarized in section 7. We
relegate the details of derivation of Green's function in the cosmic
string background to the appendix.

\section{Energy-momentum tensor: point-splitting regularization}

The energy-momentum tensor for the quantized charged massive scalar
field $\Psi(x)$ is given by expression \cite{Che,Ca} (for a review
see \cite{Ful})
\begin{eqnarray}
T^{\mu\nu}(x)=\frac 12\left[\nabla^\mu\Psi^\dag(x),\,\nabla^\nu\Psi(x)\right]_++
\frac 12\left[\nabla^\nu\Psi^\dag(x),\,\nabla^\mu\Psi(x)\right]_+ \nonumber \\
+\frac 14 g^{\mu\nu}\left[(\Box+m^2+\xi R)\Psi^\dag(x),\,\Psi(x)\right]_++
\frac 14 g^{\mu\nu}\left[\Psi^\dag(x),\,(\Box+m^2+\xi R)\Psi(x)\right]_+ \nonumber \\
+(\xi-\frac 14)g^{\mu\nu}\Box\left[\Psi^\dag(x),\,\Psi(x)\right]_+-
\xi\left(\nabla^\mu\nabla^\nu+R^{\mu\nu}\right)\left[\Psi^\dag(x),\,\Psi(x)\right]_+,\label{eq1}
\end{eqnarray}
where $\nabla_\mu$ is the covariant derivative involving both affine
and bundle connections, $\Box=\nabla_\mu\nabla^\mu$ is the covariant
d'Alembertian operator, $g_{\mu\nu}$ is the space-time metric
tensor, $R^{\mu\nu}$ is the Ricci tensor and
$R=g_{\mu\nu}R^{\mu\nu}$ is the scalar curvature of space-time,
$\xi$ is the coupling constant of the scalar field to the space-time
scalar curvature. The product of the field operators at the same
point $(x)$ is ill-defined, and this leads to the divergence of the
formal expression for the vacuum expectation value of the
energy-momentum tensor, $\langle0|T^{\mu\nu}(x)|0\rangle$. To
regularize this divergence, one uses the point splitting for the
operator product in (1):
\begin{equation}
T^{\mu\nu}(x;\,x')=D^{\mu\nu}(x;\,x')\left[T\Psi(x)\Psi^\dag(x')+T\Psi(x')\Psi^\dag(x)\right],\label{eq2}
\end{equation}
where $T$ is the symbol of chronological ordering in the operator
product and
\begin{eqnarray}
D^{\mu\nu}(x;\,x')=\left(\frac 12-\xi\right)\left(\nabla^\mu_x\nabla^\nu_{x'}+
\nabla^\nu_x\nabla^\mu_{x'}\right)\nonumber \\ +\xi g^{\mu\nu}(\Box_x+m^2+\xi R_x)+\xi g^{\mu\nu}
(\Box_{x'}+m^2+\xi R_{x'}) \nonumber \\
+2\left(\xi-\frac 14\right)g^{\mu\nu}\left(g_{\rho\sigma}\nabla_x^\rho\nabla_{x'}^\sigma-m^2-
\frac 12\xi R_x-\frac 12\xi R_{x'}\right) \nonumber \\
-\xi\left(\nabla_x^\mu\nabla_x^\nu+\nabla_{x'}^\mu\nabla_{x'}^\nu+\frac 12R_x^{\mu\nu}+
\frac 12R_{x'}^{\mu\nu}\right).\label{eq3}
\end{eqnarray}
The vacuum expectation value of tensor $T^{\mu\nu}$ (2) is well
defined. Further, if field operator $\Psi$ obeys the Klein-Gordon
equation, then the vacuum expectation value of the chronological
operator product in the right-hand side of (2) is related to Green's
function of the Klein-Gordon operator,
\begin{equation}
\langle 0|T\Psi(x)\Psi^\dag(x')|0\rangle=
-{\rm i}G(x;\,x'),\label{eq4}
\end{equation}
where
\begin{equation}
\left(\Box_x+m^2+\xi R_x\right)G(x;\,x')=\left(\Box_{x'}+m^2+\xi R_{x'}\right)G(x;\,x')=\frac{1}{\sqrt{g}}
\delta(x-x')\label{eq5}
\end{equation}
and $g=|{\rm det}g_{\mu\nu}|$. Green's function is decomposed as
\begin{equation}
G(x;\,x')=G(x;\,x')_{(S)}+G(x;\,x')_{(R)},\label{eq6}
\end{equation}
where the first term is singular, whereas the second one is regular
in the coincidence limit, $x'\rightarrow x$; the divergence of the
vacuum expectation value of tensor $T^{\mu\nu}$ (1) is due to the
contribution of $G(x;\,x')_{(S)}+G(x';\,x)_{(S)}$ to tensor
$T^{\mu\nu}$ (2). Thus, the physically meaningful (renormalized)
vacuum expectation value of the energy-momentum tensor is obtained
by taking the coincidence limit after the subtraction of the
dangerous (divergent in this limit) part:
\begin{equation}
t^{\mu\nu}(x)=-{\rm i}\lim\limits_{x'\rightarrow x}D^{\mu\nu}(x;\,x')
\left[G(x;\,x')_{(R)}+G(x';\,x)_{(R)}\right].\label{eq7}
\end{equation}
This is a quite general scheme which is only necessary but not
sufficient to define unambiguously the renormalized value. As we
shall see in the next section, to get the renormalized value in the
case of the cosmic string background, it suffices to subtract the
appropriate value in the case of the trivial (Minkowski) background.

\section{Cosmic string background: Green's function}

As is already mentioned, a cosmic string associated with spontaneous
breakdown of a gauge symmetry is characterized by a gauge field with
the strength which is directed along the string and is non vanishing
inside its core; the total flux of the strength is denoted in the
following by $\Phi$. Space-time outside the string core is the
conical space-time where a surface orthogonal to the string axis is
isometric to the surface of a cone; the deficit angle is $8\pi
G\mu$, where $G$ is the gravitational constant (squared Planck
length), $\mu$ is the tension (or energy density per unit length) of
the string, and units $c=\hbar=1$ are used. In view of the
significance of high-dimensional space-times in various aspects
\cite{Sak,Cop}, we consider a generalization of a cosmic string to a
codimension-2 brane in $(d+1)$-dimensional space-time with squared
length element
\begin{equation}
{\rm d}s^2={\rm d}t^2-{\rm d}r^2-\nu^{-2}r^2{\rm d}\varphi^2-{\rm d}{\bf z}^2_{d-2},\label{eq8}
\end{equation}
where $r$ and $\varphi$ are the polar coordinates of the conical
surface, ${\bf z}_{d-2}=(x^3,\,\ldots,\,x^d)$, $x^j$
($j=\overline{3,\,d}$) are the Cartesian coordinates of flat
($d-2$)-dimensional space, and
\begin{equation}
\nu=(1-4G\mu)^{-1}.\label{eq9}
\end{equation}
The transverse size of the cosmic string is neglected\footnote{The
limit of the transverse size tending to zero may cause problems
which were for a long discussed in the literature
\cite{Ger,Fut,Gar,Tra}. It was claimed that the source
(stress-energy tensor) of space-time (8) makes sense as a
distribution \cite{Isr,Sok}. However, a class of regular metrics can
provide for the distributional source which is concentrated on a
codimension-1, and not higher, brane \cite{Ger}. Various limiting
procedures and wider classes of metrics were considered for the case
of a codimension-2 brane, yielding an ambiguity in the relation
between the energy density per unit length of the string and the
deficit angle of space-time \cite{Ger,Fut,Gar}. Finally, a class of
metrics was found providing for the distributional source with
support on a codimension-2 brane and with the ambiguity eliminated
\cite{Tra}. This class comprises metrics which are either vanishing
(in the case of the negative deficit angle $2\pi(1-\nu^{-1})<0$) or
diverging (in the case of the positive deficit angle,
$0<2\pi(1-\nu^{-1})<2\pi$) along a normal to the brane in its
vicinity. Thus, parameter $\nu$ in (8) can take values in the range
$0<\nu<\infty$.}, and both the spatial and the bundle curvatures are
nonvanishing and singular (as distributions) in the codimension-2
brane (i.e. point in the $d=2$ case, line in the $d=3$ case, surface
in the $d=4$ case and $(d-2)$-hypersurface in the $d>4$ case). Thus,
the cosmic string background in which the scalar matter field is
quantized is characterized by two dimensionless parameters: $4G\mu$
and $e\Phi(2\pi)^{-1}$, where $e$ is the coupling constant of the
scalar to the gauge field forming the string and should not be
mistaken for the electromagnetic coupling constant.

The d'Alembertian in the cosmic string background takes form
\begin{equation}
\Box=\partial_t^2-\Delta^{({\rm tr})}-\sum\limits_{j=3}^{d}\partial_{j}^2,\label{eq10}
\end{equation}
where
\begin{equation}
\Delta^{({\rm tr})}=r^{-1}\partial_rr\partial_r+\nu^2r^{-2}\left(\partial_\varphi-
{\rm i}\frac{e\Phi}{2\pi}\right)^2\label{eq11}
\end{equation}
is the twodimensional (transverse) Laplacian and
$\partial_j=\frac{\partial}{\partial x^j}$. Green's function in the
cosmic string background obeys equations
\begin{eqnarray}
(\Box_x+m^2)G\left(t,\,r,\,\varphi,\,{\bf z}_{d-2};\,t',\,r',\,\varphi',\,{\bf z}_{d-2}'\right)\nonumber \\
=(\Box_{x'}+m^2)G\left(t,\,r,\,\varphi,\,{\bf z}_{d-2};\,t',\,r',\,\varphi',\,{\bf z}_{d-2}'\right)
\nonumber \\ =\delta(t-t')\frac{\nu}{r}\delta(r-r')\Delta(\varphi-\varphi')\delta^{(d-2)}\left(
{\bf z}_{d-2}-{\bf z}_{d-2}'\right),\label{eq12}
\end{eqnarray}
where
$\Delta(\varphi)=(2\pi)^{-1}\sum\limits_{n\in\mathbb{Z}}e^{in\varphi}$
is the delta-function for the angular variable,
$\Delta(\varphi+2\pi)=\Delta(\varphi)$, $\mathbb{Z}$ is the set of
integer numbers. It should be noted that, apart from the overall
phase factor, Green's function is periodic in the value of gauge
flux $\Phi$ with the period equal to $2\pi\,e^{-1}$, i.e. it depends
on
\begin{equation}
F=\frac{e\Phi}{2\pi}-\lshad\frac{e\Phi}{2\pi}\rshad,\label{eq13}
\end{equation}
where $\lshad u\rshad$ denotes the integer part of quantity $u$
(i.e. the integer which is less than or equal to $u$). Imposing
physically plausible conditions on Green's function at small and
large distances from the string, we solve (12) unambiguously and
obtain the following expression (see appendix):
\begin{eqnarray}
\fl G\left(t,\,r,\,\varphi,\,{\bf z}_{d-2};\,t',\,r',\,\varphi',\,{\bf z}_{d-2}'\right)=
\frac{{\rm i}m^{(d-1)/2}}{(2\pi)^{(d+1)/2}}\exp\left[\frac{{\rm i}e\Phi}{2\pi}(\varphi-\varphi')\right]
\left[\sum\limits_{l}e^{2{\rm i}lF\pi}\right.
\nonumber \\
\fl \times
\left.\frac{K_{(d-1)/2}\left(m\sqrt{-(t-t')^2+(r-r')^2+\left({\bf z}_{d-2}-{\bf z}_{d-2}'\right)^2+
4rr'\sin^2\left[(\varphi-\varphi'+2\pi l)(2\nu)^{-1}\right]}\right)}{\left\{
-(t-t')^2+(r-r')^2+\left({\bf z}_{d-2}-{\bf z}_{d-2}'\right)^2+4rr'\sin^2
[(\varphi-\varphi'+2\pi l)(2\nu)^{-1}]\right\}^{(d-1)/4}}\right. \nonumber \\
\fl \left.-\frac{\nu}{2\pi}\int\limits_{0}^{\infty}{\rm d} u\,\,\Omega_\nu^F(u,\,\varphi-\varphi')\right.\nonumber \\
\fl \left.\times\frac{K_{(d-1)/2}\left(
m\sqrt{-(t-t')^2+(r-r')^2+\left({\bf z}_{d-2}-{\bf z}_{d-2}'\right)^2+4rr'\cosh^2(u/2)}\right)}
{\left[-(t-t')^2+(r-r')^2+\left({\bf z}_{d-2}-{\bf z}_{d-2}'\right)^2+4rr'\cosh^2(u/2)\right]^{(d-1)/4}}
\right],\label{eq14}
\end{eqnarray}
where the summation is over integers $l$ that satisfy condition
\begin{equation}
\frac{\varphi'-\varphi}{2\pi}-\frac{\nu}{2}<l<\frac{\varphi'-\varphi}{2\pi}+\frac{\nu}{2},\label{eq15}
\end{equation}
$K_\alpha(u)$ is the Macdonald function of order $\alpha$, and
\begin{eqnarray}
\Omega_\nu^F(u,\,\varphi)=\frac {1}{\rm i}\left\{\frac{e^{{\rm i}(1-F)(\varphi+\nu\pi)}\cosh(F\nu u)-e^{-{\rm i}
F(\varphi+\nu\pi)}\cosh[(1-F)\nu u]}{\cosh(\nu u)-\cosh(\varphi+\nu\pi)}\right. \nonumber \\
\left.-\frac{e^{{\rm i}(1-F)(\varphi-\nu\pi)}\cosh(F\nu u)-e^{-{\rm i}F(\varphi-\nu\pi)}\cosh[(1-F)\lambda u]}
{\cosh(\nu u)-\cos(\varphi-\nu\pi)}\right\}.\label{eq16}
\end{eqnarray}

It should be emphasized that the coincidence limit, $x'\rightarrow
x$, is to be understood as limit $\varphi'\rightarrow \varphi$ taken
before the coincidence limit of at least another one variable, say,
before $r'\rightarrow r$. In this case the singularity of (14) (term
corresponding to $l=0$) at $x'\rightarrow x$ coincides with that of
Green's function in the Minkowski space-time
\begin{eqnarray}
\fl G\left.\left(t,\,r,\,\varphi,\,{\bf z}_{d-2};\,t',\,r',\,\varphi',\,{\bf z}'_{d-2}\right)\right|_{\mu=0\atop\Phi=0}
=\frac{{\rm i}m^{(d-1)/2}}{(2\pi)^{(d+1)/2}}\nonumber \\ \fl \times\frac{K_{(d-1)/2}\left(m\sqrt{-(t-t')^2+
(r-r')^2+({\bf z}_{d-2}-{\bf z}_{d-2}')^2+4rr'\sin^2[(\varphi-\varphi')/2]}\right)}
{\left\{-(t-t')^2+(r-r')^2+({\bf z}_{d-2}-{\bf z}_{d-2}')^2+4rr'\sin^2[(\varphi-\varphi')/2]\right\}^{(d-1)/4}};
\label{eq17}
\end{eqnarray}
note that the leading singularity is independent of mass:
$$
\frac{{\rm i}}{4\pi^{(d+1)/2}}\Gamma\left(\frac{d-1}{2}\right)\frac{1}{|x-x'|^{d-1}}
$$
(here $\Gamma(u)$ is the Euler gamma function). Thus, omitting the
term corresponding to $l=0$ in (14) in the coincidence limit
corresponds to subtracting (17) in the coincidence limit from (14),
and this renormalization procedure yields finite (renormalized)
vacuum expectation values.

\section{Cosmic string background: renormalized vacuum energy-momentum tensor}

The $l\neq 0$ terms in the finite sum in (14), as well as the last
term given by an integral in (14), contribute to the renormalized
vacuum expectation value of the energy-momentum tensor. Its
calculation is simplified by noting the following properties:
\begin{eqnarray}
(\partial_t+\partial_{t'})G(x;\,x')=0, \nonumber \\
(\partial_j+\partial_{j'})G(x;\,x')=0, \\
(\partial_\varphi+\partial_{\varphi'})G(x;\,x')=0. \nonumber \label{eq18}
\end{eqnarray}
In the coincidence limit we get
\begin{equation}
\lim\limits_{x'\rightarrow x}\partial_tG(x;\,x')_{(R)}=
\lim\limits_{x'\rightarrow x}\partial_jG(x;\,x')_{(R)}=0,\label{eq19}
\end{equation}
and, consequently, the following off-diagonal components of the
induced vacuum energy-momentum tensor are vanishing:
\begin{equation}
t^{01}=t^{02}=t^{0j}=t^{1j}=t^{2j}=t^{jj'}=0\,\,\,\,(j'\neq j).\label{eq20}
\end{equation}
Due to the next relation in the coincidence limit,
\begin{equation}
\lim\limits_{x'\rightarrow x}(\partial_r-\partial_{r'})G(x;\,x')_{(R)}=0,\label{eq21}
\end{equation}
the remaining off-diagonal component is vanishing as well:
\begin{equation}
t^{12}=0.\label{eq22}
\end{equation}

Further, since the Klein-Gordon equation is satisfied by Green's
function at $x\neq x'$, the following relations are valid:
\begin{equation}
\fl \left(g_{\rho\sigma}\nabla_x^\rho\nabla_{x'}^\sigma-m^2\right)G(x;\,x')
=-\left[\partial_r\partial_{r'}+\frac 12\left(r^{-1}\partial_rr\partial_r+{r'}^{-1}\partial_{r'}r'\partial_{r'}\right)
\right]G(x;\,x'),\label{eq23}
\end{equation}
\begin{equation}
\fl \nabla^\varphi\nabla^\varphi G(x;\,x')=g^{\varphi\varphi}\left(-\partial_t^2+\sum\limits_{j=3}^{d}\partial_j^2+
r^{-1}\partial_rr\partial_r-m^2\right)G(x;\,x'). \label{eq24}
\end{equation}

In view of the above, we get
\begin{equation}
D^{00}(x;\,x')=-\partial_t^2-2\left(\xi-\frac 14\right)
\left(\partial_r\partial_{r'}+r^{-1}\partial_rr\partial_r\right),\label{eq25}
\end{equation}
\begin{equation}
D^{11}(x;\,x')=\frac 12\partial_r\partial_{r'}-\frac 12\partial_r^2
+2\left(\xi-\frac 14\right)r^{-1}\partial_r,\label{eq26}
\end{equation}
\begin{equation}
\fl D^{22}(x;\,x')=\nu^2r^{-2}\left[-\partial_t^2+\sum\limits_{j=3}^{d}\partial_j^2
+2\left(\xi+\frac 14\right)r^{-1}\partial_rr\partial_r+
2\left(\xi-\frac 14\right)\partial_r\partial_{r'}-m^2\right],\label{eq27}
\end{equation}
\begin{equation}
D^{jj}(x;\,x')=-\partial_j^2+2\left(\xi-\frac 14\right)
\left(\partial_r\partial_{r'}+r^{-1}\partial_rr\partial_r\right).\label{eq28}
\end{equation}
Applying (25)-(28) to the regular part of Green's function according
to (7) and taking the coincidence limit, we obtain the renormalized
vacuum expectation value of the energy-momentum tensor:
\begin{equation}
t^{\mu'}_{\mu}=q^{\mu'}_{\mu}+s^{\mu'}_{\mu},\label{eq29}
\end{equation}
where
\begin{eqnarray}
\fl q_0^0(r)=q_j^j(r)=-\frac{4}{(4\pi)^{(d+1)/2}}\left(\frac{m}{r}\right)^{(d+1)/2}
\sum\limits_{l=1}^{\lshad\nu/2\rshad}\cos(2lF\pi)\left(\sin\frac{l\pi}{\nu}\right)^{(3-d)/2}
\Biggl\{\Biggl[\left(\sin\frac{l\pi}{\nu}\right)^{-2} \Biggr.\Biggr.\nonumber \\
\fl \Biggl.\Biggl.+2(1-4\xi)\Biggr]
K_{(d+1)/2}\left(2mr\sin\frac{l\pi}{\nu}\right)-
2(1-4\xi)mr\sin\frac{l\pi}{\nu}K_{(d+3)/2}\left(2mr\sin\frac{l\pi}{\nu}\right)\Biggr\},
\label{eq30}
\end{eqnarray}
\begin{eqnarray}
\fl q_1^1(r)=-\frac{4}{(4\pi)^{(d+1)/2}}\left(\frac{m}{r}\right)^{(d+1)/2}\sum\limits_{l=1}^{\lshad\nu/2\rshad}
\cos(2lF\pi)\left(\sin\frac{l\pi}{\nu}\right)^{(3-d)/2}\nonumber \\
\fl \times\left[\left(\sin\frac{l\pi}{\nu}\right)^{-2}-4\xi\right]K_{(d+1)/2}\left(2mr\sin\frac{l\pi}{2}\right),\label{eq31}
\end{eqnarray}
\begin{eqnarray}
\fl q_2^2(r)=-\frac{4}{(4\pi)^{(d+1)/2}}\left(\frac{m}{r}\right)^{(d+1)/2}\sum\limits_{l=1}^{\lshad\nu/2\rshad}
\cos(2lF\pi)\left(\sin\frac{l\pi}{\nu}\right)^{(3-d)/2}\nonumber \\
\fl \times\left[\left(\sin\frac{l\pi}{\nu}\right)^{-2}-4\xi\right]\left[K_{(d+1)/2}\left(2mr\sin\frac{l\pi}{2}\right)
-2mr\sin\frac{l\pi}{\nu}K_{(d+3)/2}\left(2mr\sin\frac{l\pi}{\nu}\right)\right],\label{eq32}
\end{eqnarray}
\begin{eqnarray}
\fl s^0_0(r)=s^j_j(r)=\frac{16\nu}{(4\pi)^{(d+3)/2}}\left(\frac{m}{r}\right)^{(d+1)/2}\int\limits_{1}^{\infty}
{\rm d}v\,\Lambda(v;\,F,\,\nu)\frac{v^{(3-d)/2}}{\sqrt{v^2-1}} \nonumber \\
\fl \times\left\{[v^{-2}+2(1-4\xi)]K_{(d+1)/2}(2mrv)-2(1-4\xi)mrvK_{(d+3)/2}(2mrv)\right\},\label{eq33}
\end{eqnarray}
\begin{eqnarray}
\fl s^1_1(r)=\frac{16\nu}{(4\pi)^{(d+3)/2}}\left(\frac{m}{r}\right)^{(d+1)/2}\int\limits_{1}^{\infty}
{\rm d}v\,\Lambda(v;\,F,\,\nu)\frac{v^{(3-d)/2}}{\sqrt{v^2-1}}(v^{-2}-4\xi)K_{(d+1)/2}(2mrv),\label{eq34}
\end{eqnarray}
\begin{eqnarray}
\fl s^2_2(r)=\frac{16\nu}{(4\pi)^{(d+3)/2}}\left(\frac{m}{r}\right)^{(d+1)/2}\int\limits_{1}^{\infty}
{\rm d}v\,\Lambda(v;\,F,\,\nu)\frac{v^{(3-d)/2}}{\sqrt{v^2-1}}(v^{-2}-4\xi) \nonumber \\
\fl \times \left[K_{(d+1)/2}(2mrv)-2mrvK_{(d+3)/2}(2mrv)\right],\label{eq35}
\end{eqnarray}
and
\begin{eqnarray}
\fl \Lambda(v;\,F,\,\nu)\equiv\frac 12\Omega_\nu^F(2\,{\rm arccosh}\,v,\,0)\nonumber \\
\fl =\frac{\sin(F\nu\,\pi)\cosh[2(1-F)\nu\,{\rm arccosh}v]+\sin[(1-F)\nu\,\pi]\cosh(2F\nu\,{\rm arccosh}v)}
{\cosh(2\nu\,{\rm arccosh}v)-\cos(\nu\,\pi)}; \nonumber \\ \label{eq36}
\end{eqnarray}
clearly, tensor $q_\mu^{\mu'}$ is zero for $0<\nu<2$.

One can verify that relation
\begin{equation}
\partial_rt^1_1(r)+r^{-1}\left[t^1_1(r)-t^2_2(r)\right]=0\label{eq37}
\end{equation}
is valid; hence, the induced vacuum energy-momentum tensor is
conserved:
\begin{equation}
\nabla_\mu t^{\mu}_{\mu'}=0.\label{eq38}
\end{equation}
Taking the trace of the tensor, we get
\begin{eqnarray}
\fl g_{\mu\mu'}q^{\mu\mu'}=-\frac{8}{(4\pi)^{(d+1)/2}}\left(\frac mr\right)^{(d-1)/2}
\sum\limits_{l=1}^{\lshad\nu/2\rshad}\cos(2lF\pi)
\left(\sin\frac{l\pi}{\nu}\right)^{(1-d)/2}
\nonumber \\ \fl \times\left\{4d(\xi_c-\xi)\frac mr\sin\frac{l\pi}{\nu}
\left[K_{(d+1)/2}\left(2mr\sin\frac{l\pi}{\nu}\right)\right.\right.\nonumber \\
\fl \left.\left.-mr\sin\frac{l\pi}{\nu}K_{(d+3)/2}
\left(2mr\sin\frac{l\pi}{\nu}\right)\right]-m^2K_{(d-1)/2}\left(2mr\sin\frac{l\pi}{v}\right)\right\},\label{eq39}
\end{eqnarray}
\begin{eqnarray}
\fl g_{\mu\mu'}s^{\mu\mu'}=\frac{32\nu}{(4\pi)^{(d+3)/2}}\left(\frac mr\right)^{(d-1)/2}\int\limits_{1}^{\infty}
{\rm d}v\,\Lambda(v;\,F,\,\nu)\frac{v^{(1-d)/2}}{\sqrt{v^2-1}}\left\{4d(\xi_c-\xi)\frac mrv\right.
\nonumber \\ \fl \left.\times\left[K_{(d+1)/2}(2mrv)-mrvK_{(d+3)/2}(2mrv)\right]-m^2K_{(d-1)/2}(2mrv)\right\}
,\label{eq40}
\end{eqnarray}
where
\begin{equation}
\xi_c=(d-1)(4d)^{-1}.\label{eq41}
\end{equation}
Hence, the trace is proportional to $m^2$ at $\xi=\xi_c$, and
conformal invariance is achieved in the massless limit at
$\xi=\xi_c$:
\begin{equation}
\left.\lim\limits_{m\rightarrow 0}g_{\mu\mu'}t^{\mu\mu'}\right|_{\xi=\xi_c}=0,\label{eq42}
\end{equation}
that is in accordance with a general treatment in \cite{Che,Ca}.

\section{Various limiting cases}

Using the asymptotics of the Macdonald function at large values of
its argument, we obtain the asymptotics of the induced vacuum
energy-momentum tensor at large distances from the cosmic string.
Namely, we get in the case $0<\nu\leq 2$:
\begin{equation}
t_1^1(r)=-(2mr)^{-1}\,t_2^2(r),\qquad mr\gg 1, \label{eq43}
\end{equation}
where
\begin{eqnarray}
\fl t_0^0(r)=t_j^j(r)=t_2^2(r)=-\frac{2\nu}{(4\pi)^{(d+1)/2}}(1-4\xi)\frac{\cos[(F-1/2)\nu\pi]}{\sin(\nu\pi/2)}
\left(\frac{m}{r}\right)^{(d+1)/2}\nonumber \\
\fl \times\exp(-2mr)\{1+O[(mr)^{-1}]\},\quad mr\gg 1 \label{eq44}
\end{eqnarray}
in the case $0<\nu<2$, and
\begin{eqnarray}
\fl t_0^0(r)=t_j^j(r)=t_2^2(r)=\frac{2}{(4\pi)^{d/2}}(1-4\xi)m\left(\frac mr\right)^{d/2}
\left[\cos(2F\pi)+(2F-1)\frac{\sin(2F\pi)}{\sqrt{\pi mr}}\right] \nonumber \\
\fl \times\exp(-2mr)\{1+O[(mr)^{-1}]\},\quad mr\gg 1 \label{eq45}
\end{eqnarray}
in the case $\nu=2$. We get in the case $2<\nu<\infty$:
\begin{equation}
t_1^1(r)=-\left(2mr\sin\frac \pi \nu\right)^{-1}\,t_2^2(r),\qquad mr\gg 1, \label{eq46}
\end{equation}
\begin{eqnarray}
\fl t_2^2(r)=\frac{2}{(4\pi)^{d/2}}\cos(2F\pi)\left(\sin\frac \pi\nu\right)^{-d/2}
\left[1-4\xi\left(\sin\frac\pi\nu\right)^2\right]m\left(\frac{m}{r}\right)^{d/2}\nonumber \\
\fl \times\exp\left(-2mr\sin\frac \pi\nu\right)\{1+O[(mr)^{-1}]\},\quad mr\gg 1, \label{eq47}
\end{eqnarray}
\begin{eqnarray}
 \fl t_0^0(r)=t_j^j(r)=\frac{2}{(4\pi)^{d/2}}\cos(2F\pi)\left(\sin\frac \pi\nu\right)^{2-d/2}
(1-4\xi)m\left(\frac{m}{r}\right)^{d/2}\nonumber \\
\fl \times\exp\left(-2mr\sin\frac \pi\nu\right)\{1+O[(mr)^{-1}]\},\quad mr\gg 1. \label{eq48}
\end{eqnarray}
Asymptotics (44) in the case $d=3$ was obtained earlier in
\cite{Gui}. It should be noted that the asymptotics in the case
$2<\nu<\infty$, (46)-(48), is less decreasing than that in other
cases. The asymptotics at small distances from the cosmic string
coincides with the case of the vanishing mass of the scalar field;
this case will be considered in the next section.

In the case of a global string ($\Phi=0$), we get
\begin{eqnarray}
\fl t^0_0(r)|_{F=0}=t^j_j(r)|_{F=0}=-\frac{4}{(4\pi)^{(d+1)/2}}\left(\frac{m}{r}\right)^{(d+1)/2}
\sum\limits_{l=1}^{\lshad\nu/2\rshad}\left(\sin\frac{l\pi}{\nu}\right)^{(3-d)/2}
\Biggl\{\Biggl[\left(\sin\frac{l\pi}{\nu}\right)^{-2} \Biggr.\Biggr.\nonumber \\
\fl \Biggl.\Biggl.+2(1-4\xi)\Biggr]
K_{(d+1)/2}\left(2mr\sin\frac{l\pi}{\nu}\right)-
2(1-4\xi)mr\sin\frac{l\pi}{\nu}K_{(d+3)/2}\left(2mr\sin\frac{l\pi}{\nu}\right)\Biggr\}\nonumber \\
\fl +\frac{8\nu\sin(\nu\pi)}{(4\pi)^{(d+3)/2}}\left(\frac mr\right)
^{(d+1)/2}\int\limits_{1}^{\infty}\frac{{\rm d}v}{\sqrt{v^2-1}} \nonumber \\
\fl \times\frac{v^{(3-d)/2}}{\cosh^2(\nu\,{\rm arccosh}v)-\cos^2(\nu\pi/2)}
\left\{[v^{-2}+2(1-4\xi)]K_{(d+1)/2}(2mrv)\right. \nonumber \\
\fl \left.-2(1-4\xi)mrv K_{(d+3)/2}(2mrv)\right\},\label{eq49}
\end{eqnarray}

\begin{eqnarray}
\fl t^1_1(r)|_{F=0}=-\frac{4}{(4\pi)^{(d+1)/2}}\left(\frac{m}{r}\right)^{(d+1)/2}\sum\limits_{l=1}^{\lshad\nu/2\rshad}
\left(\sin\frac{l\pi}{\nu}\right)^{(3-d)/2}\nonumber \\
\fl \times\left[\left(\sin\frac{l\pi}{\nu}\right)^{-2}-4\xi\right]K_{(d+1)/2}\left(2mr\sin\frac{l\pi}{2}\right)
\nonumber \\ \fl +\frac{8\nu\sin(\nu\pi)}{(4\pi)^{(d+3)/2}}\left(\frac{m}{r}\right)
^{(d+1)/2}\int\limits_{1}^{\infty}\frac{{\rm d}v}{\sqrt{v^2-1}} \nonumber \\
\fl \times\frac{v^{(3-d)/2}(v^{-2}-4\xi)}{\cosh^2(\nu\,{\rm arccosh}v)-\cos^2(\nu\pi/2)}
K_{(d+1)/2}(2mrv),\label{eq50}
\end{eqnarray}

\begin{eqnarray}
\fl t^2_2(r)|_{F=0}=-\frac{4}{(4\pi)^{(d+1)/2}}\left(\frac{m}{r}\right)^{(d+1)/2}\sum\limits_{l=1}^{\lshad\nu/2\rshad}
\left(\sin\frac{l\pi}{\nu}\right)^{(3-d)/2}\nonumber \\
\fl \times\left[\left(\sin\frac{l\pi}{\nu}\right)^{-2}-4\xi\right]\left[K_{(d+1)/2}\left(2mr\sin\frac{l\pi}{2}\right)
-2mr\sin\frac{l\pi}{\nu}K_{(d+3)/2}\left(2mr\sin\frac{l\pi}{\nu}\right)\right]
\nonumber \\ \fl +\frac{8\nu\sin(\nu\pi)}{(4\pi)^{(d+3)/2}}\left(\frac mr\right)
^{(d+1)/2}\int\limits_{1}^{\infty}\frac{{\rm d}v}{\sqrt{v^2-1}} \nonumber \\
\fl \times\frac{v^{(3-d)/2}(v^{-2}-4\xi)}{\cosh^2(\nu\,{\rm arccosh}
v)-\cos^2(\nu\pi/2)}
\left[K_{(d+1)/2}(2mrv)-2mrvK_{(d+3)/2}(2mrv)\right].\label{eq51}
\end{eqnarray}

In the case of the vanishing string tension ($\mu=0$) when the
string is reduced to a merely magnetic one, the result of \cite{Si3}
is recovered:
\begin{eqnarray}
\fl t^0_0(r)|_{\nu=1}=t^j_j(r)|_{\nu=1}=\frac{16\sin(F\pi)}{(4\pi)^{(d+3)/2}}\left(\frac{m}{r}\right)^{(d+1)/2}
\int\limits_{1}^{\infty}
\frac{{\rm d}v}{\sqrt{v^2-1}}\cosh\left[(2F-1){\rm arccosh}v\right] \nonumber \\
\fl \times v^{(1-d)/2}\left\{[v^{-2}+2(1-4\xi)]K_{(d+1)/2}(2mrv)-2(1-4\xi)mrvK_{(d+3)/2}(2mrv)\right\},\label{eq52}
\end{eqnarray}
\begin{eqnarray}
\fl t^1_1(r)|_{\nu=1}=\frac{16\sin(F\pi)}{(4\pi)^{(d+3)/2}}\left(\frac mr\right)^{(d+1)/2}\int\limits_{1}^{\infty}
\frac{{\rm d}v}{\sqrt{v^2-1}}\cosh[(2F-1){\rm arccosh}v] \nonumber \\
\fl \times v^{(1-d)/2}(v^{-2}-4\xi)K_{(d+1)/2}(2mrv),\label{eq53}
\end{eqnarray}
\begin{eqnarray}
\fl t^2_2(r)|_{\nu=1}=\frac{16\sin(F\pi)}{(4\pi)^{(d+3)/2}}\left(\frac{m}{r}\right)^{(d+1)/2}\int\limits_{1}^{\infty}
\frac{{\rm d}v}{\sqrt{v^2-1}}\cosh[(2F-1){\rm arccosh}v] \nonumber \\
\fl \times v^{(1-d)/2}(v^{-2}-4\xi)\left[K_{(d+1)/2}(2mrv)-2mrvK_{(d+3)/2}(2mrv)\right];\label{eq54}
\end{eqnarray}
it should be noted that temporal component (energy density) (52) at
$\xi=1/4$ was obtained in \cite{Si8,SiB}.

\section{Massless quantized matter}

Using the asymptotics of the Macdonald function at small values of
the argument, we obtain the induced vacuum energy-momentum tensor in
the case of the massless scalar field:
\begin{eqnarray}
\fl t_\mu^{\mu'}|_{m=0}=\frac{2\Gamma\left(\frac{d+1}{2}\right)}{(4\pi)^{(d+1)/2}}\,r^{-d-1}
\left\{\left[P_d(F,\,\nu)-4\xi_cP_{d-2}(F,\,\nu)\right]{\rm diag}(1,\,1,\,-d,\,1,\,\ldots,\,1)\right.\nonumber \\
\fl \left.+4(\xi-\xi_c)P_{d-2}(F,\,\nu){\rm diag}(d-1,\,-1,\,d,\,d-1,\,\ldots,\,d-1)\right\},\label{eq55}
\end{eqnarray}
where $\xi_c$ is given by (41), and
\begin{equation}
P_d(F,\,\nu)=P_d^{(q)}(F,\,\nu)+P_d^{(s)}(F,\,\nu),\label{eq56}
\end{equation}
\begin{equation}
P_d^{(q)}(F,\,\nu)=-\sum\limits_{l=1}^{\lshad \nu/2\rshad}
\cos(2lF\pi)\left(\sin\frac{l\pi}{\nu}\right)^{-d-1},\label{eq57}
\end{equation}
\begin{equation}
P_d^{(s)}(F,\,\nu)=\frac{\nu}{\pi}\int\limits_{1}^{\infty}{\rm d}v\,\frac{\Lambda(v;\,F,\,\nu)}{v^{d+1}\sqrt{v^2-1}}.\label{eq58}
\end{equation}
The tensor trace is determined by $P_{d-2}(F,\,\nu)$ only:
\begin{equation}
g_{\mu\mu'}t^{\mu\mu'}|_{m=0}=\frac{8\Gamma\left(\frac{d+1}{2}\right)}
{(4\pi)^{(d+1)/2}}r^{-d-1}(\xi-\xi_c)d(d-1)P_{d-2}(F,\,\nu).\label{eq59}
\end{equation}

In the case of the vanishing string tension, the integral in (58) is
taken, yielding (see \cite{Si3})
\begin{equation}
P_d^{(s)}(F,\,1)=\frac{\sin(F\pi)}{\pi}\,\frac{2^d\Gamma\left(\frac{d+1}{2}+F\right)\Gamma\left(
\frac{d+1}{2}+1-F\right)}{\Gamma(d+2)},\label{eq60}
\end{equation}
while $P^{(q)}_d(F,\,1)=0$.

In the case of the nonvanishing string tension, the integral in (58)
is explicitly taken for odd values of $d$ only \cite{Dow}. For this
task one should use the integral representation through contour
$C_=$ on the complex $w$-plane, see figure 2 and, e.g., (A.18) in
appendix,
\begin{equation}
P_d^{(s)}(F,\,\nu)=-\frac{\nu}{8\pi {\rm i}}\int\limits_{C_=}{\rm d}w
\frac{{\rm cosh}\left[(F-1/2)\nu w\right]}{{\rm sinh}(\nu w/2)}
[-{\rm sinh}^2(w/2)]^{-(d+1)/2}.\label{eq61}
\end{equation}
If $d$ is odd, then the last factor in (61) gives a pole at $w=0$,
and contour $C_=$ is deformed to encircle this pole, as well as
other poles on the imaginary axis; hence the integral takes form
\begin{eqnarray}
\fl P_d^{(s)}(F,\,\nu)\!=\!-P_d^{(q)}(F,\,\nu)+
\frac{1}{4\pi {\rm i}}\oint\frac{{\rm d}w}{w}\sum\limits_{n\in\mathbb{Z} \atop n\geq 0}
B_{2n}(F)\frac{(\nu w)^{2n}}{(2n)!}\left[-\sinh^2(w/2)\right]^{-(d+1)/2}\,\,({\rm odd}\,d),\nonumber \\ \label{eq62}
\end{eqnarray}
where the origin is circumvented anticlockwise, and $B_n(u)$ is the
Bernoulli polynomial of order $n$ (see, e.g., \cite{Abra}). Only the
simple pole contributes, yielding
\begin{equation}
\fl P_d(F,\,\nu)=\sum\limits_{n=0}^{(d+1)/2}C_{2n}^{d+1}\nu^{2n}B_{2n}(F)
\quad({\rm odd}\,d),\label{eq63}
\end{equation}
where
\begin{equation}
\fl C_{2n}^{d+1}=\frac{1}{(2n)!}\,\frac{1}{4\pi {\rm i}}
\oint {\rm d}w\,w^{2n-1}[-\sinh^2(w/2)]^{-(d+1)/2}.\label{eq64}
\end{equation}
The integral in (64) is in a familiar way transformed into an
integral over contour $C_=$, resulting in
\begin{eqnarray}
\fl C_{2n}^{d+1}=-\frac{1}{(2n)!}\,\frac{1}{4\pi {\rm i}}
\int\limits_{C_=}{\rm d}w\,w^{2n-1}[-\sinh^2(w/2)]^{-(d+1)/2}\nonumber \\
\fl =-\frac{1}{(2n)!}\,\frac{1}{2\pi {\rm i}}
\int\limits_{0}^{\infty} {\rm d}u\,[(u+{\rm i}\pi)^{2n-1}-(u-{\rm i}\pi)^{2n-1}]
[\cosh(u/2)]^{-d-1}.\label{eq65}
\end{eqnarray}
Although the integral in the last line of (65) can be taken with the
help of \cite{Pru},
\begin{eqnarray}
\fl C_{2n}^{d+1}=-\frac{2^d}{d!}\left\{\frac{(-1)^{n-1}}{(2n)!}\pi^{2(n-1)}\Gamma^{2}\left(\frac{d+1}{2}\right)\right.
\nonumber \\ \fl \left.+\frac{1}{n(2n-1)}\sum\limits_{l=0}^{n-2}
\frac{(-1)^l\pi^{2l}}{(2l)!}\sum\limits_{{k\in\mathbb{Z}}\atop{k\geq0}}
\frac{(-1)^k(k+d)!}{k![k+(d+1)/2]^{2(n-l)-1}}\right\},\,\,n>0,\label{eq66}
\end{eqnarray}
the obtained expression is hardly operative for $n>1$. The more
efficient expressions can be obtained for separate consecutively
decreasing values of $n$:
\begin{eqnarray}
\fl C_{d+1}^{d+1}=(-1)^{(d+1)/2}\frac{2^d}{(d+1)!}, \nonumber \\
\fl C_{d-1}^{d+1}=(-1)^{(d-1)/2}\frac{2^{d-2}}{(d-1)!}\frac{d+1}{6}, \nonumber
\\ \fl C_{d-3}^{d+1}=(-1)^{(d-3)/2}\frac{2^{d-4}}{(d-3)!}\,\frac{d+1}{72}\left(\frac 75+d\right), \\
\fl C_{d-5}^{d+1}=(-1)^{(d-5)/2}\frac{2^{d-6}}{(d-5)!}\,\frac{d+1}{432}\left[\frac{31}{35}+
\frac{7d}{5}+\frac{d(d-1)}{3}\right],\nonumber \label{eq67}
\end{eqnarray}
and so on; also, expressions in terms of the generalized Bernoulli
numbers can be used, see \cite{Dow}. In view of the above, we get
the following results for $P_d(F,\,\nu)$ (63) up to, for instance,
$d=9$:
\begin{equation}
\fl P_1(F,\,\nu)=\frac{1}{6}-\nu^2B_2(F),\label{eq68}
\end{equation}
\begin{equation}
\fl P_3(F,\,\nu)=\frac{1}{90}+\frac{\nu^4}{3}B_4(F)
+\frac 23P_1(F,\,\nu),\label{eq69}
\end{equation}
\begin{equation}
\fl P_5(F,\,\nu)=\frac{31}{9450}+\frac{\nu^4}{15}B_4(F)-\frac{2\nu^6}{45}B_6(F)
+\frac 45P_3(F,\,\nu),\label{eq70}
\end{equation}
\begin{equation}
\fl P_7(F,\,\nu)=\frac{289}{198450}+\frac{8\nu^4}{315}B_4(F)
-\frac{4\nu^6}{189}B_6(F)+\frac{\nu^8}{315}B_8(F)+\frac{6}{7}P_5(F,\,\nu),\label{eq71}
\end{equation}
\begin{eqnarray}
\fl P_9(F,\,\nu)=\frac{12637}{22453200}+\frac{19\nu^4}{3240}B_4(F)
-\frac{14\nu^6}{1215}B_6(F)+\frac{\nu^8}{405}B_8(F)-\frac{2\nu^{10}}{14175}B_{10}(F)\nonumber \\
\fl +\frac{8}{9}P_7(F,\,\nu),\label{eq72}
\end{eqnarray}
where the explicit form of relevant Bernoulli's polynomials is
\begin{eqnarray}
\fl B_2(F)=\frac{1}{6}-F(1-F),\quad B_4(F)=-\frac{1}{30}+F^2(1-F)^2,\nonumber \\
\fl B_6(F)=\frac{1}{42}-\frac 12F^2(1-F)^2-F^3(1-F)^3, \nonumber \\
\fl B_8(F)=-\frac{1}{30}+\frac{2}{3}F^2(1-F)^2+\frac{4}{3}F^3(1-F)^3+F^4(1-F)^4,  \\
\fl B_{10}(F)=\frac{5}{66}-\frac 32F^2(1-F)^2-3F^3(1-F)^3-\frac 52F^4(1-F)^4-F^5(1-F)^5.\nonumber \label{eq73}
\end{eqnarray}
Note the appearance of the quadrupled conformal coupling as a factor
in the last terms in (69)-(72) (i.e. $4\xi_c$ at $d=3,\,5,\,7$ and
9, respectively); due to this circumstance, the tensor in the
conformally invariant case ($\xi=\xi_c$) does not contain terms
which are linear in $\nu^2$ and in $F(1-F)$, only the higher integer
powers (from $\nu^4$ to $\nu^{d+1}$ and from $[F(1-F)]^2$ to
$[F(1-F)]^{(d+1)/2}$) occur. This fact holds true for arbitrary odd
values of $d$, being a consequence of relation
\begin{equation}
\fl d\,C_2^{d+1}=(d-1)C_2^{d-1},\label{eq74}
\end{equation}
which follows from the explicit expression given by (66) at $n=1$
\begin{equation}
\fl C_2^{d+1}=-\frac{\sqrt{\pi}}{2}\,\frac{\Gamma\left(\frac{d+1}{2}\right)}
{\Gamma\left(\frac{d}{2}+1\right)}.\label{eq75}
\end{equation}
As to coefficient $C_0^{d+1}$, it is positive and decreasing from
$1/6$ to $0$ as $d$ increases from $1$ to $\infty$,
\begin{equation}
\fl C_0^{d+1}=\int\limits_{0}^{\infty}\frac{{\rm d}u}{u^2+\pi^2}[\cosh(u/2)]^{-d-1}. \label{eq76}
\end{equation}
It should be noted that relations (68) and (69) are known for quite
a long time, see \cite{Fro,Dow}.

If $d$ is even, then the integrand in (61) has a branch point
singularity at $w=0$ and representation (58) is more convenient for
a numerical evaluation of $P_d^{(s)}(F,\,\nu)$ in this case.

\section{Summary}

In the present paper we have shown that a cosmic string induces the
finite energy-momentum tensor in the vacuum of the quantized massive
scalar field. Although a cosmic string in astrophysical applications
is characterized by small positive values of the deficit angle of
conical space, that corresponds to the range of parameter $\nu$ (9)
restricted to $0<1-\nu^{-1}<4\times 10^{-7}$ \cite{Bat}, our
consideration extends to the most general cases of conical space
with all possible positive and negative values of the deficit angle,
that corresponds to $0<\nu<\infty$ (see footnote \ddag). Another
generalization which is employed in the present paper is that of a
cosmic string to a codimension-2 brane in the static (ultrastatic)
locally flat $(d+1)$-dimensional space-time, see (8). The transverse
size (thickness) of the brane is neglected, whereas the mass of the
quantized scalar field is taken into account and the coupling of the
field to the scalar curvature of space-time ($\xi$) is assumed to be
arbitrary.

The general expression for components of the induced vacuum
energy-momentum tensor is given by (29)-(36). A remarkable feature
is a possibility of analytic continuation to complex values of space
dimension, that yields the tensor components as holomorphic
functions of $d$ on the whole complex $d$-plane. In the case
$0<\nu<2$, all the dependence on the brane parameters (tension and
flux) is contained in function $\Lambda$ (36), whereas the
dependence on $d$ is factored otherwise. The Aharonov-Bohm effect
\cite{Aha} is manifested in the dependence of (30)-(35) on the
fractional part of $e\Phi(2\pi)^{-1}$ (13) rather than on the flux
itself. Moreover, the tensor components are symmetric under
substitution $F\rightarrow 1-F$; consequently, the induced vacuum
energy-momentum tensor is even under charge conjugation.

A completely novel feature is that the tensor components in the case
$2\leq\nu<\infty$ attain an additional to (33)-(35) contribution
which is given by (30)-(32). This feature is analogous to the
finding of the authors of \cite{Ho, Des, Sou} that the incoming wave
in the problem of quantum-mechanical scattering on a conical
singularity is a superposition of a finite number of distorted plane
waves propagating in different (rotated) directions. To be more
precise, the number of the additional terms in the tensor
components, that equals $\lshad\nu/2\rshad$ (see (30)-(32)), is
related to the number of waves propagating out of the shadow or
double-image region in conical space, that equals
$2\lshad\nu/2\rshad+1$ (see \cite{Si5, Si10}).

Finally, it should be noted that the tensor components decrease
exponentially at large distances from the brane, see (43)-(48). If
the mass of the quantized scalar field is zero, then expressions for
the tensor components are simplified considerably, see (55)-(58). In
the case of odd dimension of space, the dependence on the flux is
contained in Bernoulli's polynomials of even order, which are
multiplied by $\nu$ in the power of the same order; we give
explicitly the results for the cases $d=3,\,5,\,7$ and 9, see
(68)-(72).

\section*{Acknowledgments}

The work was supported by the Ukrainian-Russian SFFR-RFBR project
F40.2/108 "Application of string theory and field theory methods to
nonlinear phenomena in low dimensional systems" and by special
program "Microscopic and phenomenological models of fundamental
physical processes in micro- and macroworld" of the Department of
Physics and Astronomy of the National Academy of Sciences of
Ukraine.

\renewcommand{\thesection}{A}
\renewcommand{\theequation}{\thesection.\arabic{equation}}
\setcounter{section}{1} \setcounter{equation}{0}

\section*{Appendix}

We start with the kernel of the resolvent of operator
$-\Delta^{({\rm tr})}+m^2$:
\begin{equation}
R^\omega(r,\,\varphi;\,r',\,\varphi')=\langle r,\,\varphi|(-\Delta^{({\rm tr})}+m^2-\omega^2)^{-1}|
r',\,\varphi'\rangle,\label{a1}
\end{equation}
where $\omega$ is a complex parameter. The resolvent kernel in the cosmic
string background obeys equations
\begin{equation}
\fl (-\Delta^{({\rm tr})}_x+m^2-\omega^2)R^\omega(x;\,x')=
(-\Delta^{({\rm tr})}_{x'}+m^2-\omega^2)R^\omega(x;\,x')=\frac{\nu}{r}\delta(r-r')\Delta(\varphi-\varphi'),\label{a2}
\end{equation}
where $x=(r,\,\varphi)$, $x'=(r',\,\varphi')$ and $\Delta^{({\rm
tr})}$ is given by (11). We impose the condition of regularity of
(A.1) at small distances, $r\rightarrow 0$ or $r'\rightarrow 0$,
while the condition at large distances is that (A.1) behave
asymptotically as an outgoing wave: $r^{-1/2}\exp({\rm
i}r\sqrt{\omega^2-m^2})$ at $r\rightarrow \infty$ or
$r'^{-1/2}\exp({\rm i}r'\sqrt{\omega^2-m^2})$ at $r'\rightarrow
\infty$, where a physical sheet for square root is chosen as $0<{\rm
Arg}\sqrt{\omega^2-m^2}<\pi$ \,(${\rm Im}\sqrt{\omega^2-m^2}>0$). As
a result we get
\begin{eqnarray}
\fl R^\omega(r,\,\varphi;\,r',\,\varphi')=\frac{\nu}{2\pi}\sum\limits_{n\in \mathbb{Z}}
\left\{\theta(r-r')K_{\nu[n-e\Phi(2\pi)^{-1}]}(\kappa r)I_{\nu|n-e\Phi(2\pi)^{-1}|}(\kappa r')\right.
\nonumber \\ \fl \left.+\theta(r'-r)I_{\nu|n-e\Phi(2\pi)^{-1}|}(\kappa r)K_{\nu[n-e\Phi(2\pi)^{-1}]}
(\kappa r')\right\}e^{{\rm i}n(\varphi-\varphi')},
\label{a3}
\end{eqnarray}
where $I_\alpha(u)$ and $K_\alpha(u)$ are the modified Bessel
functions of order $\alpha$ (see, e.g. \cite{Abra}), $\theta(u)$ is
the Heavyside (step) function, and
\begin{equation}
\kappa=-{\rm i}\sqrt{\omega^2-m^2},\qquad -\pi/2<{\rm Arg\,\kappa<\pi/2}\qquad ({\rm Re\,\kappa>0}).\label{a4}
\end{equation}
Choosing $r'>r$, we rewrite (A.3) as
\begin{eqnarray}
\fl R^\omega(r,\,\varphi;\,r',\,\varphi')=\frac{\nu}{2\pi}\exp\left[{\rm i}\lshad\frac{e\Phi}{2\pi}\rshad
(\varphi-\varphi')\right]\left[\sum\limits_{k\in \mathbb{Z}\atop k\geq 1}I_{\nu(k-F)}(\kappa r)
K_{\nu(k-F)}(\kappa r')e^{{{\rm i}k(\varphi-\varphi')}}\right.\nonumber \\
\fl +\left.\sum\limits_{k'\in \mathbb{Z}\atop k'\geq 0}I_{\nu(k'+F)}(\kappa r)
K_{\nu(k'+F)}(\kappa r')e^{{-{\rm i}k'(\varphi-\varphi')}}\right]=\frac{1}{4\pi}\exp\left[{\rm i}
\lshad\frac{e\Phi}{2\pi}\rshad(\varphi-\varphi')\right] \nonumber \\
\fl \times\int\limits_{0}^{\infty}\frac{{\rm d}y}{y}\exp\left(-\frac{\kappa^2rr'}{2y}-\frac{r^2+r'^2}{2rr'}y\right)
M^F_\nu(y,\,\varphi-\varphi'),
\label{a5}
\end{eqnarray}
where
\begin{equation}
M^F_\nu(y,\,\varphi)=\nu\left[\sum\limits_{k\in \mathbb{Z}\atop k\geq 1}I_{\nu(k-F)}(y)e^{{\rm i}k\varphi}+
\sum\limits_{k'\in \mathbb{Z}\atop k'\geq 0}I_{\nu(k'+F)}(y)e^{-{\rm i}k'\varphi}\right],\label{a6}
\end{equation}
$F$ is the fractional part of $e\Phi(2\pi)^{-1}$, see (13).
\begin{figure}
  \includegraphics[width=250pt]{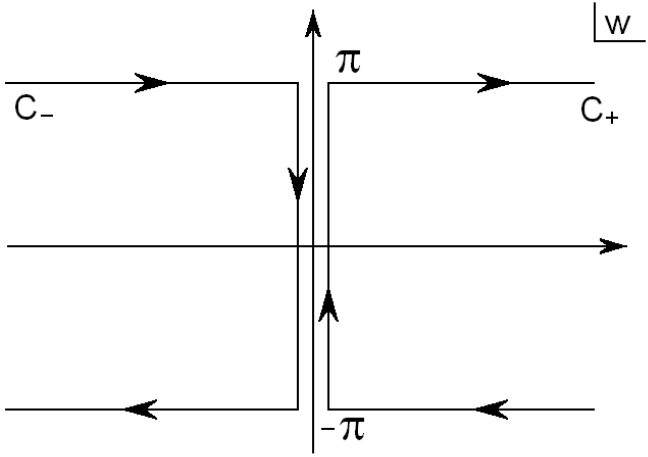}\\
  \caption{The vertical part of contour $C_+$ is infinitesimally close to the imaginary axis from the right;
  the vertical  part of contour $C_-$ is infinitesimally close to the imaginary axis from the left.}\label{1}
\end{figure}\normalsize

Using the Schl\"{a}fli contour integral representation for the
modified Bessel function
$$
I_\alpha(y)=\frac{1}{2\pi {\rm i}}\int\limits_{C_+}{\rm d}w\,e^{y\cosh w-\alpha w}=-\frac{1}{2\pi {\rm i}}
\int\limits_{C_-}{\rm d}w\,e^{y\cosh w+\alpha w},
$$
where contours $C_+$ and $C_-$ in the complex $w$-plane are depicted
in figure 1, we perform summation in (A.6) to get
\begin{eqnarray}
\fl M^F_\nu(y,\varphi)=\frac{\nu}{2\pi {\rm i}}\left[-\int\limits_{C_-}{\rm d}w\,e^{y\cosh\,w-F\nu w}
\frac{e^{\nu w+{\rm i}\varphi}}{1-e^{\nu w+{\rm i}\varphi}}+
\int\limits_{C_+}{\rm d}w\,e^{y\cosh w-F\nu w}\frac{1}{1-e^{-\nu w-{\rm i}\varphi}}\right]\nonumber \\
\fl =\frac{\nu}{2\pi i}\int\limits_{C_+ + C_-}{\rm d}w\,e^{y\cosh w-F\nu w}\frac{1}{1-e^{-\nu w-{\rm i}\varphi}}.
\label{a7}
\end{eqnarray}
Further, the union of contours $C_+$ and $C_-$ can be continuously
deformed into a contour depicted in figure 2. Taking the residues of
poles on the imaginary axis, we get
\begin{eqnarray}
M^F_\nu(y,\varphi)=\sum\limits_{l}\exp\left\{{\rm i}F(\varphi+2\pi l)+y\cos\left[(\varphi+2\pi l)\nu^{-1}\right]\right\}
\nonumber \\ -\frac{\nu e^{{\rm i}F\varphi}}{2\pi {\rm i}}\int\limits_{C_=}{\rm d} w\,e^{y\cosh\,w}\frac
{e^{(1-F)(\nu w+{\rm i}\varphi)}}{1-e^{\nu w+{\rm i}\varphi}},
\label{a8}
\end{eqnarray}
\begin{figure}
  \includegraphics[width=250pt]{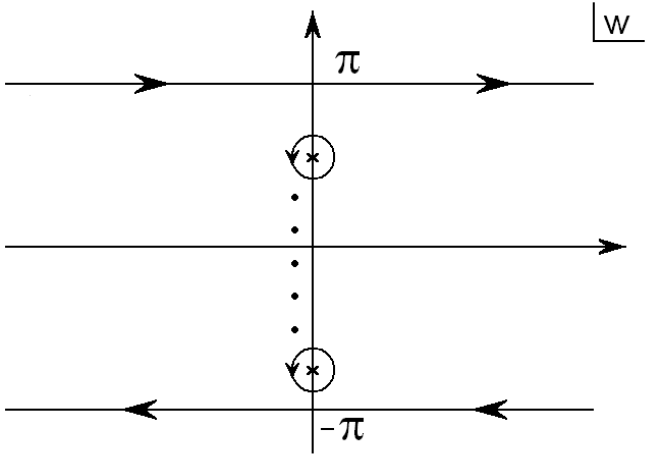}\\
  \caption{Two straight horizontal lines and circles enclosing simple poles at
  $w=-{\rm i}(\varphi+2\pi l)\nu^{-1}$ $(-\pi<{\rm Im}\,w<\pi)$.}\label{2}
\end{figure}\normalsize
where the summation is over integers $l$ satisfying condition
\begin{equation}
-\frac{\varphi}{2\pi}-\frac{\nu}{2}<l<-\frac{\varphi}{2\pi}+\frac{\nu}{2},\label{a9}
\end{equation}
and contour $C_=$ consists of two straight horizontal lines depicted
in figure 2. Substituting (A.8) into the last line of (A.5), we get
\begin{eqnarray}
\fl R^\omega(r,\varphi;\,r',\varphi')=\frac{1}{2\pi}\exp\left[{\rm i}
\frac{e\Phi}{2\pi}(\varphi-\varphi')\right] \nonumber \\
\fl \times \Biggl[\sum\limits_{l}e^{2{\rm i}lF\pi}K_0\left(\kappa\sqrt{(r-r')^2+
4rr'\sin^2[(\varphi-\varphi'+2\pi l)(2\nu)^{-1}])}\right)
\Biggr.\nonumber \\ \fl \Biggl.-\frac{\nu}{2\pi {\rm i}}\int\limits_{C_=}{\rm d} w\,\frac{e^{(1-F)[\nu w+{\rm i}
(\varphi-\varphi')]}}
{1-e^{\nu w+{\rm i}(\varphi-\varphi')}}K_0
\left(\kappa\sqrt{(r-r')^2-4rr'{\rm sinh}^2(w/2)}\right)\Biggr];
\label{a10}
\end{eqnarray}
here the summation is over integers $l$ satisfying condition (15).

With the use of the resolvent kernel one can obtain Euclidean
Green's function in 2+1-dimensional space-time with the Wick-rotated
time axis, $t=-{\rm i}\tau$:
\begin{eqnarray}
G^E(\tau,\,r,\,\varphi;\,\tau',\,r',\,\varphi')\equiv \langle\tau,\,r,\,\varphi|
\left(-\partial^2_\tau-\Delta^{({\rm tr})}+m^2\right)^{-1}|\tau',\,r',\,\varphi'\rangle\nonumber \\
=\frac{1}{2\pi}\int\limits_{-\infty}^{\infty}{\rm d} p\,e^{{\rm i}p(\tau-\tau')}R^{{\rm i}p}(r,\,\varphi;\,r',\,\varphi').
\label{a11}
\end{eqnarray}
Substituting (A.10) with $\omega={\rm i}p$ into the last line of
(A.11) and performing the integration over $p$, we get
\begin{eqnarray}
\fl G^E(\tau,\,r,\,\varphi;\,\tau',\,r',\,\varphi')=\frac{1}{2\pi^2}\exp\left[\frac{{\rm i}e\Phi}{2\pi}
(\varphi-\varphi')\right]\nonumber \\
\fl \times\left[\sum\limits_{l}e^{2{\rm i}lF\pi}\frac{\exp\left\{-m\sqrt{(\tau-\tau')^2+(r-r')^2+
4rr'\sin^2\left[(\varphi-\varphi'+2\pi l)(2\nu)^{-1}\right]}\right\}}{\sqrt{(\tau-\tau')^2+
(r-r')^2+4rr'\sin^2\left[(\varphi-\varphi'+2\pi l)(2\nu)^{-1}\right]}}\right.\nonumber \\
\fl \left.-\frac{\nu}{2\pi {\rm i}}\int\limits_{C_=}\!{\rm d} w\,\frac{e^{(1-F)[(\nu w+
{\rm i}(\varphi-\varphi')]}}{1-e^{\nu w+{\rm i}(\varphi-\varphi')}}
\frac{\exp\left[-m\sqrt{(\tau\!-\!\tau')^2+(r\!-\!r')^2\!-\!
4rr'{\rm sinh}^2(w/2)}\right]}{\sqrt{(\tau-\tau')^2+(r-r')^2-4rr'{\rm sinh}^2(w/2)}}
\right]\!.
\label{a12}
\end{eqnarray}

Euclidean Green's function in 3+1-dimensional space-time with
imaginary time $t=-{\rm i}\tau$ can be obtained as a double Fourier
transform of the resolvent kernel:
\begin{eqnarray}
\fl G^E(\tau,\,r,\,\varphi,\,z;\,\tau',\,r',\,\varphi',\,z')\equiv \langle\tau,\,r,\,\varphi,\,z|
\left(-\partial_\tau^2-\Delta^{({\rm tr})}-\partial_z^2+m^2\right)^{-1}|\tau',\,r',\,\varphi',z'\rangle \nonumber \\
\fl =\frac{1}{(2\pi)^2}\int\limits_{-\infty}^{\infty}{\rm d} q\,e^{{\rm i}q(z-z')}\int\limits_{-\infty}^{\infty}{\rm d} p\,e^{{\rm i}
p(\tau-\tau')}R^{{\rm i}\sqrt{q^2+p^2}}(r,\,\varphi;\,r',\,\varphi'),
\label{a13}
\end{eqnarray}
or equivalently:
\begin{equation}
\fl G^E(\tau,\,r,\,\varphi,\,z;\,\tau',\,r',\,\varphi',\,z')=\frac{1}{2\pi}\int\limits_{-\infty}^{\infty}
{\rm d} q\,e^{{\rm i}q(z-z')}G(\tau,\,r,\,\varphi;\,\tau',\,r',\,\varphi')|_{m\rightarrow\sqrt{m^2+q^2}}.
\label{a14}
\end{equation}
Substituting (A.12) with $m\rightarrow\sqrt{m^2+q^2}$ into (A.14),
we get
\begin{eqnarray}
\fl G^E(\tau,\,r,\,\varphi,\,z;\,\tau',\,r',\,\varphi',\,z')=\frac{m}{(2\pi)^2}\exp\left[\frac{{\rm i}e\Phi}{2\pi}
(\varphi-\varphi')\right] \nonumber \\
\fl \times\left[\sum\limits_{l}e^{2{\rm i}lF\pi}\frac{K_1\left(m\sqrt{(\tau-\tau')^2+(r-r')^2+(z-z')^2+4rr'\sin^2
[(\varphi-\varphi'+2\pi l)(2\nu)^{-1}]}\right)}{\sqrt{(\tau-\tau')^2+(r-r')^2+(z-z')^2+4rr'\sin^2
\left[(\varphi-\varphi'+2\pi l)(2\nu)^{-1}\right]}}-\right. \nonumber \\
\fl \left.-\frac{\nu}{2\pi {\rm i}}\int\limits_{C_=}{\rm d} w\,\frac{e^{(1-F)[\nu w+{\rm i}(\varphi-\varphi')}}
{1-e^{\nu w+{\rm i}(\varphi-\varphi')}}\frac{K_1\left(m\sqrt{(\tau-\tau')^2+(r-r')^2+(z-z')^2-
4rr'{\rm sinh}^2(w/2)}\right)}{\sqrt{(\tau-\tau')+(r-r')^2+(z-z')^2-4rr'{\rm sinh}^2(w/2)}}
\right].\nonumber \\
\label{a15}
\end{eqnarray}

Proceeding in this way one can get Green's function in space-time of
arbitrary dimension, since
\begin{eqnarray}
\fl G^E(\tau,\,r,\,\varphi,\,z_1,\,\ldots,\,z_{d-2};\,\tau',\,r',\,\varphi',\,z_1',\,\ldots,\,z_{d-2}')\nonumber \\
\fl =\frac{1}{2\pi}\int\limits_{-\infty}^{\infty}{\rm d} q\,e^{{\rm i}q(z_{d-2}-z_{d-2}')}G^E(\tau,\,r,\,\varphi,\,z_1,\,\ldots,\,
z_{d-3};\,\tau',\,r',\,\varphi',\,z_1',\,\ldots,\,z_{d-3}')|_{m\rightarrow \sqrt{m^2+q^2}}.\nonumber \\
\label{a16}
\end{eqnarray}
The key point is the use of relation
\begin{eqnarray}
\fl \frac{1}{2\pi}\int\limits_{-\infty}^{\infty}{\rm d} q\,e^{{\rm i} q z}\frac{(q^2+m^2)^{(d-2)/4}}{(2\pi a)^{(d-2)/2}}
K_{(d-2)/2}(a\sqrt{q^2+m^2})=\left(\frac{m}{2\pi}\right)^{(d-1)/2}\frac{K_{(d-1)/2}\left(m\sqrt{a^2+z^2}\right)}
{(a^2+z^2)^{(d-1)/4}}, \nonumber \\
\label{a17}
\end{eqnarray}
which is derived with the help of the integral representation for
the Macdonald function (see, e.g., \cite{Abra})
$$
K_\alpha(2bc)=\frac 12\left(\frac cb\right)^\alpha\int\limits_{0}^{\infty}{\rm d} y\,y^{\alpha-1}\exp\left(
-\frac{b^2}{y}-c^2y\right).
$$
Thus, Euclidean Green's function in $d+1$-dimensional space-time is
given by expression
\begin{eqnarray}
\fl G^E\left(\tau,\,r,\,\varphi,\,{\bf z}_{d-2};\,\tau',\,r',\,\varphi',\,{\bf z}_{d-2}'\right)=
\frac{m^{(d-1)/2}}{(2\pi)^{(d+1)/2}}\exp\left[\frac{{\rm i}e\Phi}{2\pi}(\varphi-\varphi')\right]
\left[\sum\limits_{l}e^{2{\rm i}lF\pi}\right.\nonumber \\
\fl \left.\times\frac{K_{(d-1)/2}\left(m\sqrt{(\tau-\tau')^2+
(r-r')^2+({\bf z}_{d-2}-{\bf z}_{d-2})^2+4rr'\sin^2[(\varphi-\varphi'+2\pi l)(2\nu)^{-1}]}\right)}
{\left\{(\tau-\tau')^2+(r-r')^2+({\bf z}_{d-2}-{\bf z}_{d-2}')^2+4rr\sin^2[(\varphi-\varphi'+2\pi l)(2\nu)^{-1}]
\right\}^{(d-1)/4}}\right.\nonumber \\
\fl \left.-\frac{\nu}{2\pi {\rm i}}\int\limits_{C_=}{\rm d}w\,
\frac{e^{(1-F)[\nu w+{\rm i}(\varphi-\varphi')]}}{1-e^{\nu w+{\rm i}(\varphi-\varphi')}}\right. \nonumber \\
\fl \left.\times\frac{K_{(d-1)/2}\left(m\sqrt{(\tau-\tau')^2+
(r-r')^2+({\bf z}_{d-2}-{\bf z}_{d-2})^2-4rr'{\rm sinh}^2(w/2)}\right)}
{\left[(\tau-\tau')^2+(r-r')^2+({\bf z}_{d-2}-{\bf z}_{d-2})^2-4rr'{\rm sinh}^2(w/2)\right]^{(d-1)/4}}\right]
.\label{a18}
\end{eqnarray}
Transforming the integral over contour $C_=$ into an integral over
the real positive semiaxis, we get
\begin{eqnarray}
\fl G^E\left(\tau,\,r,\,\varphi,\,{\bf z}_{d-2};\,\tau',\,r',\,\varphi',\,{\bf z}_{d-2}'\right)=
\frac{m^{(d-1)/2}}{(2\pi)^{(d+1)/2}}\exp\left[\frac{{\rm i}e\Phi}{2\pi}(\varphi-\varphi')\right]
\left[\sum\limits_{l}e^{2{\rm i}lF\pi}\right.\nonumber \\
\fl \left.\times\frac{K_{(d-1)/2}\left(m\sqrt{(\tau-\tau')^2+
(r-r')^2+({\bf z}_{d-2}-{\bf z}_{d-2}')^2+4rr'\sin^2[(\varphi-\varphi'+2\pi l)(2\nu)^{-1}]}\right)}
{\left\{(\tau-\tau')^2+(r-r')^2+({\bf z}_{d-2}-{\bf z}_{d-2}')^2+4rr'\sin^2[(\varphi-\varphi'+2\pi l)(2\nu)^{-1}]
\right\}^{(d-1)/4}}\right.\nonumber \\
\fl \left.-\frac{\nu}{2\pi}\int\limits_{0}^{\infty}{\rm d}u\,
\Omega_\nu^F(u,\,\varphi-\varphi')\right. \nonumber \\
\fl \left.\times\frac{K_{(d-1)/2}\left(m\sqrt{(\tau-\tau')^2+
(r-r')^2+({\bf z}_{d-2}-{\bf z}_{d-2}')^2+4rr'{\rm cosh}^2(u/2)}\right)}
{\left[(\tau-\tau')^2+(r-r')^2+({\bf z}_{d-2}-{\bf z}_{d-2}')^2+4rr'{\rm cosh}^2(u/2)\right]^{(d-1)/4}}\right]
,\label{a19}
\end{eqnarray}
where $\Omega_\nu^F(u,\,\varphi)$ is given by (16). Going over to
real time, we obtain Green's function (14).

\end{document}